\documentclass[aps,pre,superscriptaddress,groupedaddress]{revtex4}
\usepackage{graphicx}  
\usepackage{bm}        
\usepackage{amssymb}   
\usepackage{amsmath}
\usepackage{enumerate}
\usepackage{color}
\usepackage{pgfplots}

\usepackage{accents}
\MakeRobust{\underaccent} 

\begin{document}

\title{Influence of time delay on information exchanges between coupled linear stochastic systems}

\author{ M.L. Rosinberg}
\affiliation{LPTMC, Sorbonne Universit\'e, CNRS, F-75005 Paris, France}
\email{mlr@lptmc.jussieu.fr}
\author{G. Tarjus}
\affiliation{LPTMC, Sorbonne Universit\'e, CNRS,  F-75005 Paris, France}
\email{tarjus@lptmc.jussieu.fr}\author{T. Munakata}
\affiliation{Department of Applied Mathematics and Physics, 
Graduate School of Informatics, Kyoto University, Kyoto 606-8501, Japan
} 

\begin{abstract}
Time lags are ubiquitous in biophysiological processes and more generally in real-world complex networks. It has  been recently proposed to use information-theoretic tools such as transfer entropy to detect and estimate a possible delay in the couplings.  In this work, we focus on stationary  linear stochastic processes in continuous time and compute the transfer entropy in the presence of delay and correlated noises, using an approximate but numerically effective solution to the relevant Wiener-Hopf factorization problem. Our results rectify and complete the recent study of \cite{BS2017}.
\end{abstract} 

\date{\today}

\maketitle

\section{Introduction}

There is no need to overstate the prevalence of time lags in biological processes and more generally in complex networks, from neural and gene regulatory circuits to climate, traffic, communication, or computer systems, to name just a few (see e.g. \cite{N2001,A2010} and references therein).  Delays, arising from finite propagation or processing times, also play a crucial role in sensor-actuator feedback applications~\cite{AM2008,B2005}.  Particularly significant is the interplay of delays and noise which is at the origin of the complex dynamical behavior observed in a host of experimental systems~\cite{T2014}. Although these issues are increasingly the focus of theoretical and experimental investigations, there are many examples in which very little is known about the magnitude of the time lags, or even their existence, and how they are distributed in the network. As a result, this may lead to a wrong identification of the causal relationships between the various physical or chemical processes occurring in the network.

A common method in biology or climate science for estimating delays and the direction of information transfer between coupled systems is to consider temporal cross-correlation functions extracted from time-series data (see e.g. ~\cite{DCLME2008,K2014,KSL1999,GA2011}). A peak in these functions is then interpreted as the time it takes for an upstream signal  (e.g., a protein concentration)  to  influence the downstream one (e.g., a target gene). The reliability of this method is questionable, however, and the true physical meaning of the maxima in the cross correlations is often unclear~\cite{RPK2014}.
As another option, it has been recently proposed to use information-theoretic measures built on the concept of Shannon entropy and  mutual information, such as transfer entropy (TE). The idea is to identify a possible interaction delay by searching for a maximum in the TE (or some variant of it) as a function of an additional parameter, typically the prediction horizon~\cite{VWLP2011,VMKM2011,PR2011,IHHLLB2011,WPPSSLL2013}. Transfer entropy~\cite{S2000,P2001}, which is essentially a generalization of Wiener-Granger causality principle~\cite{W1956,G1969}, 
characterizes the directional information flow between two interacting random processes and  has become a popular tool for analyzing networks of interacting agents or processes, in particular in neuroscience~\cite{WRL2014,BBHL2016,CGR2018}. Whether or not this method is effective is still debated~\cite{CJJHKPP2017}, and before applying it to real data  it is worth checking the results on systems where the dynamical equations are known and that can be fully analyzed numerically or even analytically.

With this perspective in mind, the present work is motivated by a recent study~\cite{BS2017} that focuses on the determination of Granger causality (GC) from empirical sampled data produced by an underlying continuous-time process.  As a working example, the case of a bivariate linear stochastic process with delayed interaction is investigated in detail. Our objective here is not to discuss the important issue of subsampling, which is meticulously treated in \cite{BS2017} (see also \cite{ZZXC2014}) but merely to revisit the calculation of the continuous-time GC, which is equivalent to the TE in the case of multivariate Gaussian processes~\cite{BBS2009}. Indeed, it turns out that the analytical solution proposed in \cite{BS2017} for the  process with delay is incorrect when the noises acting on each  subsystem are correlated. This is not an academic issue because such correlations are often required when modeling real systems, for instance cell metabolic networks~\cite{ELSS2002, DCLME2008,K2014,HT2016,LNTRL2017}.  It is thus important to have the correct expression of the continuous-time TE before investigating the issue of delay identification (and, in a second stage, the effects of subsampling).  We also take this opportunity to rephrase the problem into a language that is perhaps more familiar to physicists, in particular those concerned with the use of information-theoretic concepts and tools in the field of stochastic and information thermodynamics~\cite{SU2012,IS2013,HBS2014,HS2014,HBS2016}.

The paper is organized as follows. In Sec. \ref{Sec:Intro}, we recall the definition of TE and its relationship with GC in the context of time-continuous stochastic processes. We also present the class of linear stochastic systems that will be considered.  
In Sec. \ref{Sec:Delay}, we then focus on a bivariate system with a time lag in one of the couplings and we describe the calculation of the TE's in  both directions. As usual with delay systems, the complication arises from the fact that the state space is infinite-dimensional.  We then introduce an approximation scheme  in the frequency domain which allows us to solve the relevant Wiener-Hopf factorization problem.   We also emphasize some points which in our opinion are not clearly stated in \cite{BS2017}, in particular the condition for the spectral expression of the TE rate to be valid. The numerical calculations presented in Sec. \ref{Sec:Numer} show that our method leads to a rapidly convergent solution, and we then investigate the issue of delay identification. A summary of the results is provided in Sec. \ref{Sec:Conclusion}. In addition, the analytical expression of an alternative, simplified version of TE is derived in the Appendix.

\section{Setup}
\label{Sec:Intro}

\subsection{Transfer entropy in continuous time}

Consider two subsystems $X_1$ and $X_2$ of a stochastic system ${\bf X}$.  The corresponding random variables or states at time $t$ are denoted by $X_1(t)$ and $X_2(t)$, respectively. As originally defined in \cite{S2000} in the discrete-time framework, the transfer entropy from $X_i$ (the ``source") to $X_j$ (the ``target")  quantifies the reduction of uncertainty in the value of $X_j(t)$ when learning the past of $X_i(t)$, if the past of $X_j(t)$ is already known.   
In continuous time, one must introduce infinitesimal increments, just as in the case of GC~\cite{CR1996}, which leads to define TE as the rate
\begin{align} 
\label{EqTE0}
{\cal T}_{i \to j}&= \lim_{h\to 0^+}\frac{1}{h}I\left[X_j(t+h):X_i^-(t)\vert X_j^-(t)\right] \, ,
\end{align}
where  $X_i^-(t)\equiv \{X_i(s): s\le t\}$ and $I$ is the conditional mutual information~\cite{CT2006}. In terms of  probability distributions, this is rewritten as
\begin{align} 
\label{EqTE1}
{\cal T}_{i \to j}&=\lim_{h\to 0^+}\frac{1}{h}\langle \ln \frac{p(X_j(t+h)\vert {\bf X}^-(t))}{p(X_j(t+h)\vert X_j^-(t))}\rangle\, ,
\end{align}
where ${\bf X}^-(t)\equiv (X_1^-(t),X_2^-(t))$.
Since we only focus on stationary processes, ${\cal T}_{i\to j}$ is here independent of $t$. Note that the {\it whole} past history of both the source and the target up to time $t$ are taken into account in Eqs. (\ref{EqTE0}) and (\ref{EqTE1}). Conditioning the mutual information on $X_j^-(t)$ is natural because the marginal -or ``coarse-grained"- processes $X_1$ and $X_2$ are generally non-Markovian even when the joint process ${\bf X}$ is Markovian. But it is also sensible to take into account the whole vector $X_i^-(t)$ and not only the latest state $X_i(t)$, as is done in other context~\cite{HBS2016,MS2017}.  This seems particularly justified for the class of non-Markovian processes that are studied in the following [cf. Eqs. (\ref{EqDyn})-(\ref{EqDelay})].   (In the original definition of TE in discrete time~\cite{S2000,P2001}, the lengths of the two state vectors $X_1^-(t)$ and $X_2^-(t)$ -i.e. the number of time bins in the past- are generally finite and possibly different. This definition can also be extended to continuous time~\cite{SLP2016}.)

Although the rate ${\cal T}_{i \to j}$ is an interesting quantity {\it per se}, for instance in the context of stochastic thermodynamics~\cite{HBS2016}, it cannot be used to infer a possible time lag in the couplings. Just as in the discrete-time framework~\cite{WPPSSLL2013}, this role is devoted to the finite-horizon TE  
\begin{align} 
\label{EqTijh}
T_{i \to j}(h)&=I\left[X_j(t+h):X_i^-(t)\vert X_j^-(t)\right] \nonumber\\
&=\langle \ln \frac{p(X_j(t+h)\vert {\bf X}^-(t))}{p(X_j(t+h)\vert X_j^-(t))}\rangle
\end{align}
which in the terminology of forecasting~\cite{L2005} quantifies how much the prediction of $X_j(t+h)$ is improved by using both $X_i^-(t)$ and $X_j^-(t)$ rather than $X_j^-(t)$ alone. This is clearly quite similar to the definition of the continuous-time finite-horizon GC in \cite{BS2017}, which itself extends the classic discrete-time definition~\cite{L2005,DR1998}. (Note in passing that $T_{i \to j}(h)$, as a relative entropy, is a non-negative quantity and vanishes at $h=0$ by construction.) The main difference is that TE is model-independent whereas GC is commonly defined in the context of vector autoregressive processes (VAR). However, TE and GC become fully equivalent  when the variables are Gaussian distributed, with a simple factor of $1/2$ relating the two quantities~\cite{BBS2009}. Indeed, since the entropy of  Gaussian distributions is directly expressed in terms of their covariance matrix~\cite{CT2006}, Eq. (\ref{EqTijh}) then yields
\begin{align} 
\label{EqTijh1}
T_{i \to j}(h)&=\frac{1}{2}\ln\frac{\sigma'_{jj}(h)}{\sigma_{jj}(h)}\, ,
\end{align}
where
\begin{align} 
\label{Eqsigmajj}
\sigma_{jj}(h)=\langle[X_j(t+h)-\langle X_j(t+h)\vert {\bf X}^-(t)\rangle]^2\rangle
\end{align}
and
\begin{align} 
\label{Eqsigmajjp}
\sigma'_{jj}(h)&=\langle [X_j(t+h)-\langle X_j(t+h)\vert X_j^-(t)\rangle]^2\rangle 
\end{align}
are the variances of $p(X_j(t+h)\vert {\bf X}^-(t))$ and $p(X_j(t+h)\vert X_j^-(t))$, respectively.  Note that we have assumed that the two variables $X_1(t)$ and $X_2(t)$ are univariate, which will be the situation considered hereafter (see \cite{BBS2009} for the generalization to multivariate variables).
In the language of forecasting, $\langle X_j(t+h)\vert {\bf X}^-(t)\rangle $ and  $\langle X_j(t+h)\vert X_j^-(t)\rangle $ are interpreted as the minimum mean-square error (MMSE) estimates of $X_j(t+h)$ and $\sigma_{jj}(h)$ and $\sigma'_{jj}(h)$ are the corresponding mean-square prediction errors. The present work is mainly concerned with the calculation of these quantities in the presence of delayed interactions.

Notwithstanding the valuable arguments for conditioning $X_j(t+h)$ on the infinite past histories of $X_1(t)$ and $X_2(t)$, it is also useful from a practical viewpoint to  consider a simplified version of $T_{i \to j}(h)$ that only involves the states at time $t$, 
\begin{align} 
\label{EqTbarh}
{\overline T}_{i \to j}(h)&= I\left[X_j(t+h):X_i(t)\vert X_j(t)\right] \nonumber\\
&=\langle \ln \frac{p(X_j(t+h)\vert {\bf X}(t))}{p(X_j(t+h)\vert X_j(t))}\rangle\, .
\end{align}
In the case of Gaussian distributed variables, ${\overline T}_{i \to j}(h)$ is given by an expression similar to Eq. (\ref{EqTijh1}) whose explicit calculation is presented in Appendix A. It it worth noticing that ${\overline T}_{i \to j}(h)$ is an upper bound on $T_{i \to j}(h)$ if the joint process ${\bf X}$ is Markov~\cite{note00} (and in turn, ${\overline {\cal T}}_{i \to j}$,  the slope at the origin, is an upper bound on ${\cal T}_{i \to j}$~\cite{HBS2014,HBS2016}). However, this is no longer true in the general case.

\subsection{Class of models}

In \cite{BS2017}, the following class of linear stochastic integro-differential equation was  introduced 
\begin{align} 
\label{EqDyn}
\dot {\bf X}(t)=-\int_0^{\infty} ds\:{\bf A}(s){\bf X}(t-s)+{\boldsymbol \xi}(t)\, ,
\end{align}
where ${\bf X}$ is an $n$-dimensional vector process, ${\bf A}(s)$ is an $n\times n$ matrix of functions or generalized functions (distributions), and  ${\boldsymbol \xi}(t)$ is an $n$-dimensional vector of (generally correlated) Gaussian white noises. Eq. (\ref{EqDyn}) is viewed as the continuous-time analog of a VAR representation (see \cite{CR1996} for mathematical details), and this type of equation, which can be obtained through the linearization of  nonlinear problems, appears in various research fields where the history of the state variables must be taken into account, e.g. in econometry, biology, or control theory.   Depending on the context, the time lags may then be either discrete or distributed according to some density function. This latter case often occurs in the modeling of biological processes~\cite{M1989,C2010,H2011}. In the following, we shall instead focus on the case of discrete delays, so that Eq. (\ref{EqDyn}) takes the form of a linear stochastic differential delay equation, 
\begin{align} 
\label{EqDyn1}
\dot {\bf X}(t)=-\sum_{\alpha=1}^N {\bf A}_{\alpha}{\bf X}(t-\tau_{\alpha})+{\boldsymbol \xi}(t)\,,
\end{align}
with possibly $N$ distinct delays $\tau_{\alpha}$~\cite{GMKC2014}. In recent years, such multivariate, multi-delayed equations have been used to study synchronization problems in complex networks (see e.g. \cite{HSK2012}). However, for simplicity, and given the purpose of this work, we will only introduce a single delay $\tau$ in one of the couplings and consider a bivariate system, as already stated.

\section{Bivariate linear process with a time-delayed coupling}
\label{Sec:Delay}

For definiteness, let us assume that the delay takes place in the feedback from $X_2$ to $X_1$. Eq. (\ref{EqDyn1}) then becomes
\begin{align}
\label{EqDelay}
\dot {\bf X}(t)=-\left(
\begin{array}{cc}
a_{11}&0 \\
a_{21}&a_{22}
\end{array}\right)
{\bf X}(t)-\left(
\begin{array}{cc}
 0&a_{12} \\  0&0
 \end{array}
\right){\bf X}(t-\tau) +{\boldsymbol \xi}(t)\, . 
\end{align}
where $\xi_1(t)$ and $\xi_2(t)$ are zero-mean Gaussian white noises with covariances $\langle \xi_i(t)\xi_j(t')\rangle=2D_{ij}\delta(t-t')$. We stress that we do not assume independent noises as is usually done in the context of  stochastic thermodynamics (the so-called bipartite assumption~\cite{SU2012,IS2013,HBS2014,HS2014,HBS2016}). It is clear that the case of a time lag in the coupling from $X_1$ to $X_2$ follows by exchanging the labels $1$ and $2$.  On the other hand,  the two directions are not equivalent for a given model, and for the process described by Eq. (\ref{EqDelay}) we will see that  the computation of the TE in the direction $2\to 1$  is significantly more difficult than in the direction $1\to 2$. One should also keep in mind that time-delayed interactions typically lead to bifurcations and complicated dynamics~\cite{N2001,A2010}. This is an interesting issue in itself, but to simplify the forthcoming discussion we assume that the delay and the coupling parameters $a_{ij}$  are such that a stable stationary solution exists (in other words, the spectral density matrix is bounded for all values of the frequency $\omega$).  Moreover, in Sec. \ref{subsec:T21}, to further simplify the model, we will completely suppress the possible occurrence of instabilities  by setting $a_{21}=0$, which corresponds to model studied in section 4 of \cite{BS2017}.

Since we only focus on the stationary regime we can assume that the process  started at $t_0=-\infty$ and forget about the initial condition. The solution of Eq. (\ref{EqDelay}) then reads
\begin{align} 
\label{EqH}
{\bf X}(t)=\int_{-\infty}^t ds\:{\bf H}(t-s){\boldsymbol \xi}(s)\, ,
\end{align}
where ${\bf H}(t)$ is the response (or Green's or transfer) functions matrix.
Equivalently, in Fourier space or frequency domain,
\begin{align} 
{\bf X}(\omega)={\bf H}(\omega){\boldsymbol \xi}(\omega)\, ,
\end{align}
where 
\begin{align}
\label{EqHomega}
{\bf H}(\omega)&\equiv \int_{-\infty}^{+\infty} dt\: e^{i\omega t}{\bf H}(t)=\frac{1}{(a_{11}-i\omega)(a_{22}-i\omega)-a_{12}a_{21}e^{i\omega \tau}}
\left(
\begin{array}{cc}
  a_{22}-i\omega&-a_{12}e^{i\omega \tau}\\
  -a_{21}&a_{11}-i\omega
\end{array}
\right)\, .
\end{align}
The power-spectrum matrix whose elements are the Fourier transform of the stationary time-dependent correlation functions $\phi_{ij}(t)=\langle X_i(t')X_j(t'+t)\rangle$ is then given by 
\begin{align} 
\label{EqSpectrum}
{\bf S}(\omega)={\bf H}(\omega)(2{\bf D}) {\boldsymbol H}^{*}(\omega)\, ,
\end{align}
where $2{\bf D}$ is the diffusion matrix with elements $2D_{ij}$ and the subscript $*$ denotes complex conjugate and matrix transpose.

\subsection{Transfer entropy in the direction $1 \to 2$}
\label{subsec:T12}

\subsubsection{Finite-horizon TE}

We begin with the calculation of the TE in the direction $1 \to 2$ which is fairly straightforward.  Although part of the material in this section may be viewed as a mere application to the bivariate case of the formalism presented in \cite{BS2017}  (with Granger causality replaced by transfer entropy), it is included to keep the paper self-contained. This is also a useful preparation for the calculations of Sec. \ref{subsec:T21}.

The starting point is  Eq. (\ref{EqTijh1}) with $i=1,j=2$, which requires to compute $\sigma_{22}(h)$ and $\sigma'_{22}(h)$, and thus the associated MMSE's. The essential ingredient for computing these quantities is to have a one-to-one correspondence between the stationary process or subprocess under consideration and the corresponding forcing white noise(s). In other words, the  process or subprocess must be invertible (or {\it minimum-phase} in the language of control theory~\cite{AM2008,B2005}): Fixing the trajectory of the process or subprocess up to time $t$ must be equivalent to fixing the trajectory of the  noise(s) and vice versa. 

When the conditioning involves the past of the joint process ${\bf X}$ (represented either by Eq. (\ref{EqDelay}) or Eq. (\ref{EqH}) which are the continuous-time analogues of the vector autoregressive and moving average representations~\cite{L2005}),  the calculation of the MMSE is immediate. Starting from 
\begin{align} 
X_2(t+h)=\int_{-\infty}^{t+h}ds\: [H_{21}(t+h-s)\xi_1(s)+H_{22}(t+h-s)\xi_2(s)]\, ,
\end{align}
we readily obtain 
\begin{align} 
\langle X_2(t+h)\vert {\bf X}^-(t)\rangle=\int_{-\infty}^{t}ds\: [H_{21}(t+h-s)\xi_1(s)+H_{22}(t+h-s)\xi_2(s)]\, ,
\end{align}
since the noises are fixed for $s\le t$ by Eq. (\ref{EqDelay}) and average to zero in the time interval $[t,t+h]$.
Eq. (\ref{Eqsigmajj}) then yields 
\begin{align} 
\label{Eqsigma22}
\sigma_{22}(h)&=2\int_0^h dt\:[D_{11}H_{21}^2(t)+D_{22}H_{22}^2(t)+2D_{12}H_{22}(t)H_{21}(t)]\, .
\end{align}

The calculation of $\langle X_2(t+h)\vert X_2^-(t)\rangle$ is less straightforward because fixing the marginal process $X_2$  alone does not fix the noises $\xi_1$ and $\xi_2$. Instead, one must find a coarse-grained representation of $X_2$ similar to Eq. (\ref{EqDelay}),
\begin{align} 
\label{Eqvar2}
\dot X_2(t)=-\int_0^{\infty} ds\:A'_{22}(s)X_2(t-s)+\xi'_2(t)\, ,
\end{align}
where $A'_{22}(s)$ is a kernel to be determined and $\xi'_2(t)$ is a Gaussian white noise, for instance with the same variance $2D_{22}$ as $\xi_2(t)$.
Then, starting from  the equation
\begin{align} 
\label{EqMVA2}
X_2(t)=\int_{-\infty}^{t}ds\: H'_{22}(t-s)\xi'_2(s)\, ,
\end{align}
where $H'_{22}(t)$ is the ``inverse" of $A'_{22}(t)$ [see below Eq. (\ref{EqA2p0})], and using the same reasoning as above, we obtain
\begin{align} 
\label{EqX2h}
\langle X_2(t+h)\vert X_2^-(t)\rangle=\int_{-\infty}^t ds\: H'_{22}(t+h-s)\xi'_2(s)
\end{align}
and in turn 
\begin{align} 
\label{Eqsigma22p}
\sigma'_{22}(h)=2D_{22}\int_0^h dt \: H_{22}^{'2}(t)\, .
\end{align}

The response function $H'_{22}(t)$ must be causal and is easily found by going to Fourier space. Indeed, Eq. (\ref{EqMVA2}) implies that the power spectral density (PSD) $S_{22}(\omega)=\langle X_2(\omega)X_2(-\omega)\rangle$ is given by
\begin{align} 
S_{22}(\omega)=2D_{22}\vert H'_{22}(\omega)\vert^2\, .
\end{align}
On the other hand,  Eq. (\ref{EqSpectrum}) tells us that 
\begin{align} 
S_{22}(\omega)&= 2D_{22}\vert H_{22}(\omega)\vert^2+2D_{11}\vert H_{21}(\omega)\vert^2+ 2D_{12}[H_{22}(\omega)H_{21}(-\omega)+H_{22}(-\omega)H_{21}(\omega)] \, ,
\end{align}
which is conveniently rewritten as
\begin{align} 
\label{PSD22}
S_{22}(\omega)= 2D_{22}\vert H_{22}(\omega)\vert^2 \frac{\omega^2+r_2^2}{\omega^2+a_{11}^2}\, ,
\end{align}
where  
 \begin{align} 
\label{Eqr2}
r_2=\sqrt{a_{11}^2+ \frac{D_{11}}{D_{22}}a_{21}^2-2 \frac{D_{12}}{D_{22}}a_{11}a_{21}}\, .
\end{align}
Since $H_{22}(\omega)$ is the Fourier transform of a causal function, the Wiener-Hopf factorization of $S_{22}(\omega)$ is simple and gives  
\begin{align} 
\label{EqHpdelay}
H'_{22}(\omega)=H_{22}(\omega)\frac{r_2 -i\omega}{a_{11}-i\omega}= \frac{r_2 -i\omega}{(a_{11}-i\omega)(a_{22}-i\omega)-a_{12}a_{21}e^{i\omega \tau}}\, .
\end{align}
(In turn, one can readily check that the noise defined by Eq. (\ref{EqMVA2}) and given in Fourier space by $\xi'(\omega)=(-a_{21}\xi_1(\omega)+(a_{11}-i\omega)\xi_2(\omega))/(r_2-i\omega)$ is indeed white.) By construction, $H'_{22}(\omega)$ has no poles in the upper half of the complex plane and since we have chosen $r_2>0$ in Eq. (\ref{Eqr2}), it is also zero-free in this region. The minimum-phase condition -a prerequisite for Eq. (\ref{EqX2h})- is thus satisfied.

For a given choice of the model parameters, the response functions $H_{ij}(t)$ and $H'_{22}(t)$ can be computed numerically by taking the corresponding inverse Fourier transforms, and $\sigma_{22}(h)$ and $\sigma'_{22}(h)$ are then obtained from Eqs (\ref{Eqsigma22}) and (\ref{Eqsigma22p}). For brevity, we do not present a numerical study here.
On the other hand, it is instructive to look at the explicit representation of the marginal process $X_2$ provided by Eq. (\ref{Eqvar2}). By construction, the Fourier transform of the kernel $A'_{22}(t)$ is obtained as
\begin{align} 
\label{EqA2p0}
A'_{22}(\omega)&\equiv\frac{1}{H'_{22}(\omega)}+i\omega\, ,
\end{align}
which yields
\begin{align} 
A'_{22}(\omega)&=\frac{(a_{11}-i\omega)(a_{22}-i\omega)}{r_2-i\omega}-a_{12}a_{21}\frac{e^{i\omega \tau}}{r_2-i\omega}\nonumber\\
&=a_{11}+a_{22}-r_2+\frac{(r_2-a_{11})(r_2-a_{22})}{r_2-i\omega}-a_{12}a_{21}\frac{e^{i\omega \tau}}{r_2-i\omega}\, .
\end{align}
As a result,
\begin{align} 
\label{EqA2p}
A'_{22}(t)= (a_{11}+a_{22}-r_2)\delta(t)+(r_2-a_{11})(r_2-a_{22})e^{-r_2t}\Theta(t)-a_{12}a_{21}e^{-r_2(t-\tau)}\Theta(t-\tau) \, ,
\end{align}
and Eq. (\ref{Eqvar2})  reads
\begin{align} 
\label{Eqx2delay}
\dot X_2(t)&= -(a_{11}+a_{22} -r_2) X_2(t)-(r_2-a_{11})(r_2-a_{22})\int_{-\infty}^t ds\: e^{-r_2(t-s)}X_2(s)\nonumber\\
&+a_{12}a_{21}\int_{-\infty}^t ds\:e^{-r_2(t-s)}X_2(s-\tau) +\xi'_2(t)\, .
\end{align}
Finally, by splitting the integrals into two parts, $\int_{-\infty}^t ds=\int_{-\infty}^{t-\tau}ds+\int_{t-\tau}^t ds$, and performing some simple manipulations, we can transform the equation into
\begin{align} 
\label{Eqx2delay1}
\dot X_2(t)&=-(a_{11}+a_{22}-r_2) X_2(t)-[(r_2-a_{11})(r_2-a_{22})-a_{12}a_{21}e^{r_2\tau}]\int_{-\infty}^{t-\tau} ds\:e^{-r_2(t-s)}X_2(s)\nonumber\\
&-(r_2-a_{11})(r_2-a_{22})\int_{t-\tau}^t ds\:e^{-r_2(t-s)}X_2(s)+\xi'_2(t)\, .
\end{align}
This is an interesting representation of the coarse-grained dynamics of $X_2$ because it shows that a significant simplification occurs if the delay $\tau$ satisfies the condition
\begin{align} 
\label{Eqdelay}
a_{12}a_{21}e^{r_2 \tau}=(r_2-a_{11})(r_2-a_{22})\, .
\end{align}
The second term in the r.h.s. of Eq. (\ref{Eqx2delay1}) then vanishes, and although the dynamics is still non-Markovian, the dependence on the past is now limited to a finite time interval of duration $\tau$.

\subsubsection{TE rate}

By definition, the TE rate ${\cal T}_{1 \to 2}$ is  the slope of $T_{1 \to 2}(h)$ at $h=0^+$. After expanding $\sigma_{22}(h)$ and $\sigma'_{22}(h)$  in powers of $h$  and using $H_{21}(0^+)=0$ and $H_{22}(0^+)=H'_{22}(0^+)=1$, we obtain
\begin{align} 
T_{1 \to 2}(h)=\frac{1}{2}\ln \frac{2D_{22}[h+\dot H'_{22}(0^+)h^2+{\cal O}(h^3)]}{2D_{22}h+2[D_{22}\dot H_{22}(0^+)+D_{12}\dot H_{21}(0^+)]h^2+{\cal O}(h^3)}\, ,
\end{align}
and then
\begin{align} 
\label{EqT1hderiv}
{\cal T}_{1 \to 2}&=\frac{1}{2}[\dot H'_{22}(0^+)-\dot H_{22}(0^+)-\frac{D_{12}}{D_{22}}\dot H_{21}(0^+)]\, ,
\end{align}
which is the two-dimensional version of Eq. (61) in \cite{BS2017} (with the usual multiplicative factor $1/2$ coming from the replacement of  GC by the corresponding TE). In order to obtain the explicit expressions of $\dot H_{21}(0^+)$ and $\dot H_{22}(0^+)$, we then use  the equation
\begin{align} 
\label{EqHdot}
\dot {\bf H}(t)&=-\int_0^t ds\: {\bf A}(s) {\bf H}(t-s)\, ,\ \ t\ge 0\, ,
\end{align}
which is obtained by differentiating Eq. (\ref{EqH}) with respect to $t$ and identifying with Eq. (\ref{EqDelay}) (see e.g. Appendix F in ~\cite{BS2017}). Specifically,
\begin{align}
\label{EqHdot1}
\dot {\bf H}(t)=-\left(
\begin{array}{cc}
  a_{11}&0\\
  a_{21}&a_{22}
\end{array}
\right) {\bf H}(t)-\left(
\begin{array}{cc}
  0&a_{12}\\
  0&0
\end{array}
\right) {\bf H}(t-\tau)\Theta(t-\tau)\, .
\end{align}
Together with the condition ${\bf H}(0^+)={\bf I}$, where ${\bf I}$ is the unity matrix, this readily yields $\dot H_{21}(0)=-a_{21}$ and $\dot H_{22}(0)=-a_{22}$. 
Likewise, $\dot H'_{22}(0^+)$ is obtained from the equation
\begin{align} 
\label{EqH2dot}
\dot H'_{22}(t)&=-\int_0^t ds\: A'_{22}(s) H'_{22}(t-s)\, ,
\end{align}
with $A'_{22}(t)$  given by Eq. (\ref{EqA2p}). Expressly,
\begin{align} 
\dot H'_{22}(t)&=-(a_{11}+a_{22}-r_2)H'_{22}(t)-(r_2-a_{11})(r_2-a_{22})\int_0^tds\:e^{-r_2(t-s)}H'_{22}(s)\nonumber\\
&+a_{12}a_{21}e^{r_2\tau}\Theta (t-\tau)\int_{0}^{t-\tau}ds\: e^{-r_2(t-s)}H'_{22}(s) \, ,
\end{align} 
from which we find that $\dot H'_{22}(0^+)=r_2-a_{11}-a_{22}$.
Inserting these expressions of $\dot H_{21}(0^+)$, $\dot H_{22}(0^+)$ and $\dot H'_{22}(0^+)$ into Eq. (\ref{EqT1hderiv}), we finally obtain
\begin{align} 
\label{EqTE12new}
{\cal T}_{1 \to 2}&=\frac{1}{2}[r_2-a_{11}+\frac{D_{12}}{D_{22}}a_{21}]\, .
\end{align}
Therefore the TE rate in the direction $1\to 2$ does not depend on $\tau$, a result that was not obvious from the outset because of the bidirectional character of the coupling between the two sub-processes. As a matter of fact,  ${\overline {\cal T}}_{1 \to 2}$, the simplified version of the TE rate that only takes into account the information provided by the states at time $t$ and whose expression is given by Eq. (\ref{EqoverT12final}) in Appendix A, does depend on $\tau$.

\subsubsection{Spectral expression of the TE rate}
\label{subsec:Spectral}

As originally introduced in the context of  VAR processes~\cite{G1982}, there is a spectral version of GC that is used, especially in neuroscience~\cite{FAS2013}, to analyze causal relationships in the frequency domain. The continuous-time version is briefly presented in \cite{BS2017}, but the conditions for the validity of this spectral representation are not discussed. This will play a important role in Sec. \ref{subsec:T21}, and for completeness we revisit the derivation, focusing again on TE instead of GC.

It is instructive to first consider the case  $D_{12}=0$ (i.e., the joint process $\bf X$ is bipartite). The PSD $S_{22}(\omega)$ then reduces to two terms, 
\begin{align} 
S_{22}(\omega)= 2D_{22}\vert H_{22}(\omega)\vert^2+2D_{11}\vert H_{21}(\omega)\vert^2\, .
\end{align}
The first one can be viewed as the intrinsic contribution of the subprocess $X_2$ to its (auto) spectrum whereas the second one can be viewed as the causal part due to $X_1$. Following \cite{G1982}, this suggests to adopt the quantity 
\begin{align} 
\label{Eqt12}
t_{1\to 2}(\omega)&\equiv \frac{1}{2}\ln \frac{S_{22}(\omega)}{2D_{22}\vert H_{22}(\omega)\vert^2}
\end{align} 
as a measure of the transfer entropy from $X_1$ to $X_2$ in the frequency domain~\cite{C2011}. However, two requirements must be fulfilled: i) $t_{1\to 2}(\omega)$ must be a non-negative quantity and ii) the TE rate in the time domain must be the average of the spectral TE  over all frequencies, i.e.,
\begin{align} 
\label{Eqspectral1}
{\cal T}_{1 \to 2}&=\frac{1}{2}\int_{-\infty}^{\infty} \frac{d\omega}{2\pi}\ln \frac{S_{22}(\omega)}{2D_{22}\vert H_{22}(\omega)\vert^2}\, .
\end{align}
The first condition is obviously fulfilled, and to check the second one we replace $H_{22}(\omega)$ and $S_{22}(\omega)$ by their expressions, Eqs. (\ref{EqHomega}) and (\ref{PSD22}) respectively, and integrate over $\omega$. This gives
\begin{align} 
\int_{-\infty}^{\infty} \frac{d\omega}{2\pi}\:t_{1\to 2}(\omega)&=\frac{1}{2}\int_{-\infty}^{\infty} \frac{d\omega}{2\pi}\:\ln \frac{r_2^2+\omega^2}{a_{11}^2+\omega^2}=\frac{1}{2}(r_2-\vert a_{11}\vert)\, .
\end{align}
which indeed coincides with Eq. (\ref{EqTE12new}) when $D_{12}=0$, {\it but at the condition that $a_{11}>0$}~\cite{note7}.

Following again \cite{G1982} and the literature on GC~\cite{DCB2006,BBS2010}, Eq. (\ref{Eqspectral1}) can be generalized to the case of correlated noises ($D_{12}\ne 0$). This is done by performing a linear transformation ${\widetilde {\boldsymbol \xi}}(t)={\bf U}{\boldsymbol \xi}(t)$ that makes the covariance matrix of the transformed noises diagonal. Specifically, by choosing 
\begin{align}
{\bf U}=\left(
\begin{array}{cc}
1  &-\frac{D_{12}}{D_{22}}  \\
  0&1
\end{array}
\right) \ ,
\end{align}
we get
\begin{align}
2{\widetilde {\bf D}}=2{\bf U}{\bf D}{\bf U}^T=2\left(
\begin{array}{cc}
D_{11}-\frac{D_{12}^2}{D_{22}}  & 0\\
  0&D_{22}
\end{array}
\right) \, ,
\end{align}
and the dynamics of the transformed vector ${\widetilde{\bf X}}(t)={\bf U}{\bf x}(t)$ is now governed by the  equation 
$\dot {\widetilde{\bf X}}(t)=-\int_0^{\infty}\widetilde{{\bf A}}(s){\widetilde{\bf X}}(t-s)+{\widetilde {\boldsymbol \xi}}(t)$ with ${\widetilde {\bf A}}={\bf U}{\bf A}{\bf U}^{-1}$. Likewise,  
\begin{align}
{\widetilde {\bf H}}(\omega)={\bf U}{\bf H}(\omega){\bf U}^{-1}=\left(
\begin{array}{cc}
H_{11}-\frac{D_{12}}{D_{22}}H_{21}   & H_{12}+\frac{D_{12}}{D_{22}}(H_{11}-H_{22}) -(\frac{D_{12}}{D_{22}})^2H_{21} \\
  H_{21}&H_{22}+\frac{D_{12}}{D_{22}}H_{21}
\end{array}
\right) \, ,
\end{align}
and 
\begin{align}
{\widetilde {\bf S}}(\omega)={\bf U}{\bf S}(\omega){\bf U}^T=\left(
\begin{array}{cc}
S_{11}-2\frac{D_{12}}{D_{22}}S_{12}+ (\frac{D_{12}}{D_{22}})^2 S_{22} &S_{12}-\frac{D_{12}}{D_{22}}S_{22}\\
  S_{12} -\frac{D_{12}}{D_{22}}S_{22}&S_{22}
\end{array}
\right) \, ,
\end{align}
where the dependence of the functions $H_{ij}$ and $S_{ij}$ on $\omega$ is dropped for brevity. The crucial feature is that the TE rate ${\cal T}_{1 \to 2}$  is invariant under the linear transformation defined by the matrix ${\bf U}$. Indeed, since $\widetilde{D}_{12}=0$,  we have from Eq. (\ref{EqT1hderiv})
\begin{align} 
\label{EqtildeT12}
{\widetilde {\cal T}}_{1 \to 2}&=\frac{1}{2}[\dot {\widetilde H}'_2(0^+)-\dot {\widetilde H}_{22}(0^+)]\nonumber\\
&=\frac{1}{2}[\dot H'_{22}(0^+)-\dot H_{22}(0^+)-\frac{D_{12}}{D_{22}}\dot H_{21}(0^+)]\nonumber\\
&={\cal T}_{1 \to 2}\, ,
\end{align}
where we have used the fact that $\widetilde{S}_{22}(\omega)=S_{22}(\omega)$ implies ${\widetilde H}'_2(\omega)=H'_{22}(\omega)$. Accordingly, by applying the spectral decomposition  (\ref{Eqspectral1}) to the transformed variables $\widetilde X_1$ and $\widetilde X_2$ and  going back to the original variables, we obtain
\begin{align} 
\label{Eqspectral2}
{\cal T}_{1 \to 2}&=\frac{1}{2}\int_{-\infty}^{\infty} \frac{d\omega}{2\pi}\ln \frac{\widetilde{S}_{22}(\omega)}{2{\widetilde D}_{22}\vert {\widetilde H}_{22}(\omega)\vert^2}\nonumber\\
&=\frac{1}{2}\int_{-\infty}^{\infty} \frac{d\omega}{2\pi}\ln \frac{S_{22}(\omega)}{2D_{22}\vert H_{22}(\omega)+(D_{12}/D_{22})H_{21}(\omega)\vert^2}\, .
\end{align}

This expression (with the labels $1$ and $2$ exchanged) will play an important role in Sec. \ref{subsec:T21} as it will give a closed-form expression of the TE rate ${\cal T}_{2 \to 1}$. However, there is a serious caveat. Replacing $H_{22}(\omega)$, $H_{21}(\omega)$, and $S_{22}(\omega)$ by their expressions and integrating over $\omega$, we find that  Eq. (\ref{EqTE12new}) is recovered {\it only if  the condition ${\widetilde a}_{11}=a_{11}-(D_{12}/D_{22})a_{21}>0$ is satisfied}. This of course generalizes the condition $a_{11}>0$ that was found  in the case $D_{12}=0$. Otherwise, Eq. (\ref{Eqspectral2}) underestimates the actual value of the TE rate in the time domain, as was already pointed out in \cite{G1982} for the discrete-time GC (see also footnote 6 in \cite{BS2017}). 

What is the rationale for the condition ${\widetilde a}_{11}>0$?  Since ${\widetilde H}_{22}(\omega)=(\tilde a_{11}-i\omega)/[(\tilde a_{11}-i\omega)(\tilde a_{22}-i\omega)-\tilde a_{12}\tilde a_{21}e^{i\omega \tau}]$ (cf. Eq. (\ref{EqHomega}) with $a_{ij}$ replaced by $\widetilde a_{ij}$), this condition guarantees that ${\widetilde H}_{22}(\omega)$ has no zeros in the upper half of the complex $\omega$-plane (it has no poles in this region since ${\widetilde H}_{22}(t)$ is causal). To summarize, the condition for the spectral expression to be valid is that the stationary process ${\widetilde X}_2(\omega)={\widetilde H}_{22}(\omega)\xi_2(\omega)$ is minimum-phase. There is no reason for this condition to be always satisfied for a time-delayed process governed by Eq. (\ref{EqDelay}), not to mention the more general Eq. (\ref{EqDyn}), and it must be carefully checked on a case-by-case  basis.

\subsection{Transfer entropy in the direction $2 \to 1$}
\label{subsec:T21}

We now turn to the calculation of the TE in the direction $2 \to 1$ and to simplify the forthcoming analysis we set the  parameter $a_{21}$ to zero from the outset. This makes the coupling unidirectional, and $X_2(t)$ becomes a simple Ornstein-Uhlenbeck process that drives $X_1(t)$ at a fixed delay $\tau$. This is the model introduced in section 4 of Ref.~\cite{BS2017}, which is regarded as the ``minimal" continuous-time version of a VAR process.
Interestingly, this also corresponds to the model of a cellular signaling pathway considered in \cite{HSWIH2016}, in which $X_1(t)$ and $X_2(t)$ represent the deviations from the mean of active kinase populations (these quantities can be treated as continuous variables by assuming a chemical Langevin description~\cite{note5}). 
Of course, the fact that $X_2$ is now an autonomous process implies that $T_{1\to 2}(h)$ and thus ${\cal T}_{1\to 2}$ vanish identically. Moreover,  the stationary state is  stable for all values of $\tau$. 

 Following \cite{BS2017}, we set $a_{11}=a>0,a_{22}=b>0, a_{12}=-c$ and we assume that the two noises $\xi_1$ and $\xi_2$ have the same variance $2D_{11}=2D_{22}=1$ to further restrict the parameter space. The parameter $\rho=2D_{12}$  (with $-1<\rho <1$) quantifies the correlation between the noises.  The response functions for $t\ge 0$ now have  very simple expressions in the time domain, 
\begin{align} 
\label{EqHtdelay}
H_{11}(t)&=e^{-at}\nonumber\\
H_{12}(t)&=c\:\frac{e^{-a(t-\tau)}-e^{-b(t-\tau)}}{b-a}\Theta(t-\tau)\nonumber\\
H_{21}(t)&=0\nonumber\\
H_{22}(t)&=e^{-bt}\, , 
\end{align}
and  $T_{2 \to 1}(h)$  is  obtained from Eq. (\ref{EqTijh1}) with $i=2, j=1$ where $\sigma_{11}(h)$ and $\sigma'_{11}(h)$ are the variances of $p(X_1(t+h)\vert {\bf X}^-(t))$ and $p(X_1(t+h)\vert X_1^-(t))$, respectively. Likewise, Eq. (\ref{EqT1hderiv}) is replaced by 
\begin{align} 
\label{EqT21hderiv}
{\cal T}_{2\to 1}&=\lim_{h\to 0^+}\frac{1}{h}T_{2 \to 1}(h)\nonumber\\
&=\frac{1}{2}[\dot H'_{11}(0^+)-\dot H_{11}(0^+)-\frac{D_{12}}{D_{11}}\dot H_{12}(0^+)]\nonumber\\
&=\frac{1}{2}[\dot H'_{11}(0^+)+a]\, ,
\end{align}
as $\dot H_{11}(0^+)=-a$ and $\dot H_{12}(0^+)=0$ from Eq. (\ref{EqHdot1}). 

\subsubsection{Wiener-Hopf factorization}

It should be clear from the previous section that the main task is to compute the response function $H'_{11}(t)$. This requires the factorization of the PSD $S_{11}(\omega)=\langle X_1(\omega)X_1(-\omega)\rangle$ which reads
\begin{align} 
\label{EqS11BS}
S_{11}(\omega)= \frac{\omega^2+b^2+c^2+2\rho c \:v(\omega,b,\tau)}{(a^2+\omega^2)(b^2+\omega^2)}\, ,
\end{align}
where $v(\omega,b,\tau)\equiv b\cos\omega \tau-\omega \sin\omega \tau$ (for comparison we use the same notations as \cite{BS2017}). This turns out to be a nontrivial operation. In \cite{BS2017}, it is claimed that  the causal factor $H'_{11}(\omega)$ satisfying $S_{11}(\omega)= \vert H'_{11}(\omega)\vert^2$ is given by  
\begin{align} 
\label{EqHpBS}
H'_{11}(\omega)= \frac{\sqrt{(1-\rho^2)c^2+(b+\rho c\cos\omega \tau)^2}-i(\omega-\rho c \sin \omega \tau)}{(a-i\omega)(b-i\omega)}\, . 
\end{align}
However, this statement is wrong when $\rho\ne 0$ because the inverse Fourier transform of this function is not causal. This can be readily seen by setting $\omega=x+iy$ and considering the limit $y\to +\infty$, which yields
\begin{align} 
H'_{11}(\omega)\sim \frac{\rho c}{2y^2}(\sqrt{e^{-2i\tau x}}-e^{-i\tau x})e^{\tau y}+{\cal O}(1/y) \, .
\end{align}
Hence,  $H'_{11}(\omega)$ diverges like $-\rho c y^{-2}e^{-i\tau x}e^{\tau y}$ if $\cos \tau x<0$ and the condition for applying Jordan's lemma is not satisfied. Accordingly, the  inverse Fourier transform does not vanish for $t<0$, as can be checked numerically. Moreover, $H_1'(t=0^+)$ is not equal to $1$, contrary to what  it should be (see Fig. \ref{Fig5} below), which implies that $\sigma'_1(h)=2D_{11}\int_0^hdt\: H'_1(t)^2\ne 2D_{11}h[1+{\cal O}(h)]$ so that the formula ${\cal T}_{2\to 1}=\lim_{h\to 0^+}(2h)^{-1}\ln [\sigma'_{11}(h)/\sigma_{11}(h)]$ gives an infinite result. This is of course a serious shortcoming.

Before presenting our solution to the factorization problem, let us explain why this operation is nontrivial, even from the numerical point of view. 
First, one could try to apply the standard Wiener-Hopf method~\cite{N1988} and transform the multiplicative factorization problem into an additive one by taking the logarithm of  $S_{11}(\omega)$. In order to have a function that goes to $1$ as $\vert \omega \vert \to \infty$, one may consider the ratio $K(\omega)=S_{11}(\omega)/S_{11}(\omega,\rho=0)=[\omega^2+b^2+c^2+2\rho c \:v(\omega,b,\tau)]/[\omega^2+b^2+c^2]$, and $H'_{11}(\omega)$ is then obtained as 
\begin{align} 
H'_{11}(\omega)=\frac{\sqrt{b^2+c^2}-i\omega}{(a-i\omega)(b-i\omega)}K_+(\omega)\, ,
\end{align}
where $K_+(\omega)$ is the causal factor of $K(\omega)$ given by
\begin{align} 
K_+(\omega)&=\exp\Big[\frac{1}{2i\pi} \int_{i\delta-\infty}^{i\delta+\infty} d\zeta\:\frac{\ln K(\zeta)}{\zeta-\omega}\Big]\, .
\end{align}
In this formula, $\omega$ must lie above $\delta$ and the integration path must belong to a finite-width strip ${\cal D}$ around the real axis where $K(\omega)$ is analytic and free of zeros. The problem with this procedure is that the numerator of $K(\omega$) [i.e., the function $\omega^2+b^2+c^2+2\rho c \:v(\omega,b,\tau)$] has infinitely many zeros in the complex $\omega$-plane when $\rho\ne 0$. Since there does not seem to be any simple and systematic way of computing these zeros for arbitrary values of the parameters, determining the zero-free strip ${\cal D}$ is a daunting task.

Alternatively, one could try to solve the  problem directly in the time domain. Recall that in order to compute $T_{2 \to 1}(h)$,  we need to calculate  the MMSE estimate $\langle X_1(t+h)\vert X_1^-(t)\rangle$, which is the orthogonal projection of $X_1(t+h)$ onto the trajectory $X_1^-(t)$. It thus satisfies the equation
\begin{align} 
\label{Eqprojection2}
\langle [X_1(t+h)-\langle X_1(t+h)\vert X_1^-(t)\rangle]\vert X_1(s)\rangle =0\: \: \: \: \forall \: s\le t\, ,
\end{align}
and is a linear functional of $X_1^-(t)$,
\begin{align}
\label{Eqprojection3} 
\langle X_1(t+h)\vert X_1^-(t)\rangle=  \int_{-\infty}^t ds\: f_h(t-s)X_1(s)\, ,
\end{align}
where $f_h(t)$ is an unknown function to be determined from Eq. (\ref{Eqprojection2}). (To be precise, the kernel $f_h(t)$ must also include a term proportional to the Dirac distribution $\delta(t)$ which singles out the dependence on $X_1(t)$.) Inserting Eq. (\ref{Eqprojection3}) into Eq. (\ref{Eqprojection2}) and changing  variables  yields the Wiener-Hopf  integral equation
\begin{align} 
\label{EqWH1}
\phi_{11}(t+h)=\int_{0}^{\infty} ds\: \phi_{11}(t-s) f_h(s)\: \: \: \: \forall \: t\ge 0\, ,
\end{align}
where $\phi_{11}(t)\equiv \langle X_1(0)X_1(t)\rangle$  is the inverse Fourier transform of  $S_{11}(\omega)$. Since $\phi_{11}(t)$ is a combination of exponentials [see Eqs. (78) and (79) in \cite{BS2017},  where $\phi_{11}(t)$ is denoted $\Gamma_{xx}(t)$], one could hope to find some systematic procedure to solve Eq. (\ref{EqWH1}) and determine $f_h(t)$, at least numerically. However, this goal cannot be achieved because $\phi_{11}(t)$ has different expressions for $t<\tau$ and $t>\tau$: $f_h(t)$ is then an infinite sum of functions defined in the successive intervals $[0,\tau], [\tau,2\tau],[2\tau,3\tau]$, etc., with  the function in the interval $n\tau\le t\le (n+1)\tau$ depending on the function in the next interval.  Therefore, a  ``step-by-step" solution of Eq. (\ref{EqWH1}) is impossible.

Our solution to the factorization problem consists in replacing the delay term $e^{i\omega \tau}$ in the frequency domain by an ``all-pass" (i.e., with unit amplitude) rational function of the form ${\cal Q}_n(-\omega)/{\cal Q}_n(\omega)$, where ${\cal Q}_n(\omega)$ is a polynomial with no zeros in the upper half of the complex $\omega$-plane. This is a classic procedure in the field of control systems~\cite{N2001}, and several choices of ${\cal Q}_n$ are possible, in particular Pad\'e approximants (which are also often used  to approximate Wiener-Hopf kernels~\cite{A2000}). After various trials, we have found that the simplest and yet effective approximation for the problem at hand is the so-called {\it Laguerre shift} formula
\begin{align}
\label{EqLaguerre}
e^{i\omega \tau}\approx \frac{(1+\frac{i\omega \tau}{2n})^n}{(1-\frac{i\omega \tau}{2n})^n}
\end{align}
which introduces a single pole of multiplicity $n$ at $\omega=-i2n/\tau$.  The convergence rate of this approximation for $n \to \infty$ has been studied in detail in the literature~\cite{L1994,MP1999} and the formula is successfully used in robust control as it is easy to implement with analog filters.   Accordingly, the expression (\ref{EqS11BS}) of $S_{11}(\omega)$ is now replaced by 
\begin{align} 
\label{EqS11new}
S_{11,n}(\omega)=\frac{{\cal P}_n(\omega)}{(a^2+\omega^2)(b^2+\omega^2)(1+\frac{\omega^2\tau^2}{4n^2})^n}\, ,
\end{align}
where 
\begin{align} 
{\cal P}_n(\omega)=(\omega^2+b^2+c^2)(1+\frac{\omega^2\tau^2}{4n^2})^n+\rho c [(1+i\frac{\omega \tau}{2n})^{2n}(b+i\omega)+(1-i\frac{\omega \tau}{2n})^{2n}(b-i\omega)]
\end{align}
is an even polynomial of order $2n+2$. The factorization problem now boils down to finding all the roots of a polynomial, a standard numerical task. Since ${\cal P}_n(\omega)$ has no real roots~\cite{note10}, Eq. (\ref{EqS11new}) can be rewritten as
\begin{align} 
S_{11,n}(\omega)=\frac{\prod_{k=1}^{n+1}(\omega-\omega_k)(\omega-\omega_k^*)}{(a^2+\omega^2)(b^2+\omega^2)(4n^2/\tau^2+\omega^2)^n}\, ,
\end{align}
where $\omega_k$ denotes a root with a negative imaginary part. The causal factor $H'_{11,n}(\omega)$ is then readily obtained as  
\begin{align} 
\label{EqH1new}
H'_{11,n}(\omega)= \frac{i\prod_{k=1}^{n+1}(\omega-\omega_k)}{(\omega+ia)(\omega+ib)(\omega+2in/\tau)^{n}}\, ,
\end{align}
where the factor $i$ is included in order that $H'_1(t=0^+)=\lim_{\omega \to \infty}-i\omega H'_{11}(\omega)=1$, as it must be. From this we can compute numerically the inverse Fourier transform $H'_{11,n}(t)$ and then
\begin{align}
\label{EqTn21}
T^{(n)}_{2 \to 1}(h)\equiv \frac{1}{2}\ln \frac{\sigma'_{11,n}(h)}{\sigma_{11,n}(h)}\, ,
\end{align}
where 
\begin{align}
\sigma'_{1,n}(h)=\int_0^h dt \: H^{'2}_{11,n}(t)\, ,
\end{align} 
\begin{align}
\label{EqSigma11}
\sigma_{11,n}(h)=\int_0^h[H_{11}^2(t)+H_{12,n}^2(t)+2\rho H_{11}(t)H_{12,n}(t)]\, ,
\end{align}
and $H_{12,n}(t)$ is the inverse Fourier transform of $H_{12,n}(\omega)=(1+\frac{i\omega \tau}{2n})^n/[(1-\frac{i\omega \tau}{2n})^n(a-i\omega)(b-i\omega)]$. 

Remarkably, this procedure leads to an explicit and concise expression of the TE rate ${\cal T}_{2 \to 1}^{(n)}$ in terms of the roots $\omega_k$. Similarly to Eq. (\ref{EqT21hderiv}), one has
\begin{align} 
{\cal T}_{2 \to 1}^{(n)}&=\lim_{h\to 0^+}\frac{1}{h}T^{(n)}_{2 \to 1}(h)\nonumber\\
&=\frac{1}{2}[\dot H'_{11,n}(0^+)-\dot H_{11}(0^+)-\frac{D_{12}}{D_{11}}\dot H_{12,n}(0^+)]\, ,
\end{align}
where $H'_{11,n}(t)$ for $t\ge 0$ is obtained by using Cauchy's residue theorem  as
\begin{align} 
\label{EqH1newt}
H'_{11,n}(t)=A_n e^{-at}+B_ne^{-bt}+Q_n(t)e^{-2nt/\tau} \, ,
\end{align}
with $A_n=\prod_{k=1}^{n+1}(a-i\omega_k)/[(a-b)(a-2n/\tau)^n]$, $B_n=\prod_{k=1}^{n+1}(b-i\omega_k)/[(b-a)(b-2n/\tau)^n]$, and $Q_n(t)= q_n^{(0)}+q_n^{(1)}t+...q_n^{(n-1)}t^{n-1}$.  Moreover, it can be shown that $\dot H'_{12,n}(0^+)=(-1)^nc$. As a result,
\begin{align} 
\label{EqT21n}
{\cal T}_{2 \to 1}^{(n)}&=\frac{1}{2}[-aA_n-bB_n+(q_n^{(1)}-\frac{2n}{\tau}q_n^{(0)})+a+(-1)^{n+1}\rho c]\, .
\end{align}
Using $q_n^{(0)}=1-A_n-B_n$ (as $H'_{11,n}(0^+)=1$) and $q_n^{(1)}=-\lim_{\omega \to \infty} \omega^2[H'_{11,n}(\omega)-A_n/(a-i\omega)-B_n/(b-i\omega)-q_n^{(0)}/(2n/\tau-i\omega)]$, we find that the coefficients of $A_n$ and $B_n$  cancel, and after some algebra we finally obtain
\begin{align} 
\label{EqT21exact1}
{\cal T}_{2 \to 1}^{(n)} &=\frac{1}{2} [a+(-1)^{n+1}\rho c-\frac{2n}{\tau}-\lim_{\omega \to \infty} \omega^2\big(H'_{11,n}(\omega)-\frac{1}{2n/\tau-i\omega}\big)]\nonumber\\
&=\frac{1}{2}[-b+(-1)^{n+1}\rho c+i\sum_{k=1}^{n+1}\omega_k-\frac{2n^2}{\tau}]\, ,
\end{align}
which is the explicit expression  announced above. Note that $ {\cal T}_{2 \to 1}^{(n)}$ and thus ${\cal T}_{2 \to 1}=\lim_{n \to \infty} {\cal T}_{2 \to 1}^{(n)}$ are invariant in the change $(\rho,c)\to (-\rho,-c)$ and do not depend on $a^{-1}$, the  intrinsic relaxation time of the process $X_1$. This is not the case for ${\overline {\cal T}}_{2 \to 1}$, whose expression is given by Eq. (\ref{EqoverT21final})  in  Appendix A.  

Of course, this is still formal and a problem of practicality remains: How large must $n$ be to provide an accurate estimation of ${\cal T}_{2 \to 1}$ and, more generally, of $T_{2 \to 1}(h)$? Although we have no rigorous mathematical answer to this question~\cite{note1}, numerical calculations show that the convergence to the asymptotic limit is quite fast (see Figs. \ref{Fig3} and \ref{Fig5} below). As shown in Appendix B, this is not the case when  the delay kernel $\delta(t-\tau)$ is approximated in the time domain by a sequence of gamma distributions~\cite{M1989,C2010,H2011}.  In addition, we have another way to assess the accuracy of Eq. (\ref{EqT21exact1})  which is to compare with the predictions of the spectral formula for ${\cal T}_{2 \to 1}$. The latter is indeed exact in a certain range of the  parameters, as we now discuss.

\subsubsection{Spectral expression of the TE rate}
 \label{subsec:Spectral2}

 The spectral expression of ${\cal T}_{2 \to 1}$ is obtained by exchanging  the labels $1$ and $2$ in Eq. (\ref{Eqspectral2}). For the present simplified model, one has
\begin{align}
\label{Eqnewprocess}
\widetilde H_{11}(\omega)=H_{11}(\omega)+\rho H_{12}(\omega)=\frac{b-i\omega +\rho ce^{i\omega \tau}}{(a-i\omega)(b-i\omega)}\, ,
\end{align}
which gives (cf. Eq. (83) in \cite{BS2017})
\begin{align}
\label{EqSpec}
{\cal T}_{2 \to 1}&=\frac{1}{2}\int_{-\infty}^{\infty}\frac{d\omega}{2\pi}\ln \frac{\omega^2+ b^2+c^2+2\rho cv(\omega,b,\tau)}{\omega^2+ b^2+\rho^2c^2+2\rho cv(\omega,b,\tau)}\, .
\end{align}
However, as was stressed above in Sec. \ref{subsec:Spectral}, the domain of validity of the spectral expression is limited and must be carefully determined, a task that has been overlooked in \cite{BS2017}. Otherwise,  the actual TE rate is underestimated.  According to the previous discussion, the correct result is obtained if $\widetilde H_{11}(\omega)$ has no zeros in the upper half-plane. The  stationary process ${\widetilde X}_1(\omega)=\widetilde H_{11}(\omega) \xi_1(\omega)$ is then minimum-phase, like all stochastic processes considered in this work. 

Remarkably, the solution to this problem is already available in the literature. Indeed, it turns out that the equation $f(\omega)\equiv b-i\omega +\rho c e^{i\omega \tau}=0$ that determines the zeros of $\widetilde H_{11}(\omega)$ is also the characteristic equation that determines the stability of the linear equation 
\begin{align}
\label{EqSDDE}
\dot x(t)=-bx(t)-\rho cx(t-\tau)
\end{align}
which has been widely studied in the literature on delay differential equations. For instance, Ref.~\cite{CG1982} (see also \cite{KM1992} and Theorem 8.6 in \cite{C2010}) tells us that this equation is asymptotically stable when all the roots of $f(\omega)$ have a negative real part, a condition that is always satisfied if $-b< \rho c\le b$ and always violated if $\rho c<-b$, whatever the value of $\tau$ (recall that $-1<\rho<1$ and $b>0$ in the present model). In the case $\rho c >b$, there is a critical value of the delay $\tau^*=\arccos[-b/(\rho c)]/\sqrt{\rho^2c^2-b^2}$ beyond which the condition is violated (when $\tau=\tau^*$, a Hopf bifurcation occurs). Note that this has nothing to do with the stability of the joint process ${\bf X}$ itself: As we have already mentioned, the stationary state is stable for all values of $\tau$. 

Likewise, there is  a spectral representation of the nth-order approximant, given by
\begin{align}
\label{EqT21n1}
{\cal T}_{2 \to 1}^{(n)}=\frac{1}{2}\int_{-\infty}^{\infty}\frac{d\omega}{2\pi}\ln \frac{{\cal P}_n(\omega)}{{\cal P}_n(\omega)-c^2(1-\rho^2)[1+\omega^2 \tau^2/(4n^2)]^n}\, ,
\end{align}
whose  domain of validity depends on $n$.

\section{Numerical illustration}
\label{Sec:Numer}

\subsection{Convergence with $n$}

\begin{figure}[hbt]
\begin{center}
\includegraphics[width=7cm]{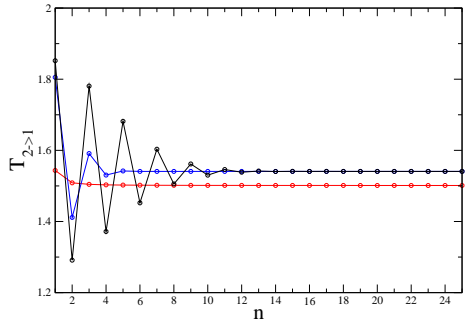}
 \caption{\label{Fig1} (Color on line)  Evolution of ${\cal T}_{2 \to 1}^{(n)}$ computed from Eq. (\ref{EqT21exact1}) as a function of $n$ for $\rho=0.2$, $c=4$, and three values of the delay: $\tau=0.5$ (red symbols), $\tau=5$ (blue symbols), $\tau=30$ (black symbols) ($b^{-1}$ is taken as the time unit). The exact asymptotic results given by Eq. (\ref{EqSpec})  are $1.501$ for $\tau=0.5$ and  and $1.541$ for $\tau=5$ and $Ö\tau=30$.}
\end{center}
\end{figure}

We first consider the issue of the convergence of ${\cal T}_{2 \to 1}^{(n)}$  with $n$. A typical example is shown in Fig. \ref{Fig1} where $\rho$ and $c$ are chosen such that the spectral formula (\ref{EqSpec}) gives the exact  value of  ${\cal T}_{2 \to 1}$ for all values of $\tau$. As announced, the good news is that ${\cal T}_{2 \to 1}^{(n)}$ converges quite rapidly towards the exact asymptotic value, even when $\tau$ is much larger than  the relaxation time $b^{-1}$ of the process $X_2$.   (In this figure and in the following we take $b^{-1}$ as the time unit.) For instance,  the relative error between ${\cal T}_{2 \to 1}^{(n)}$ and ${\cal T}_{2 \to 1}$ for $\tau=30$ is already  less that $1$\% with $n=10$. Therefore, there is no need to use large values of $n$, which would have been a practical limitation to our solution of the  Wiener-Hopf factorization~\cite{note0}.
\begin{figure}[hbt]
\begin{center}
\includegraphics[width=7cm]{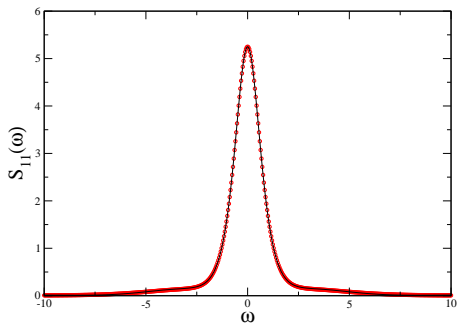}
 \caption{\label{Fig2} (Color on line)  Comparison between the exact PSD $S_{11}(\omega)$ given by Eq. (\ref{EqS11BS}) (solid line) and the approximate PSD  $S_{11}^{(n)}(\omega)=\vert H'_{1,n}\vert^2$  with $n=4$ (red circles) for $\rho=0.5$ and $\tau=1$.  The model parameters are $a=2,b=1,c=4$.}
\end{center}
\end{figure}
Note that  the exact PSD $S_{11}(\omega)$ is very well reproduced by $S_{11,n}=\vert H'_{11,n}(\omega)\vert^2$ even for $n$ small, as shown in  Fig. \ref{Fig2} where $n=4$. However, it is well-known in the field of  Wiener-Hopf factorization (see e.g. \cite{A2000}) that by itself this is not a good criterion for assessing the accuracy of the factorization: For instance, the inverse Fourier transform of the function $H'_{11}(\omega)$ given by Eq. (\ref{EqHpBS}) [which, by construction, exactly reproduces $S_{11}(\omega)$] is neither causal nor equal to $1$ at $t=0$, which implies that the corresponding TE rate is infinite, as we have already pointed out. This is illustrated in Fig. \ref{Fig3} which also shows the evolution of $H'_{11,n}(t)$ with $n$. The small wiggles for  $t<\tau$ are a consequence of the Laguerre formula  (\ref{EqLaguerre}) and they become negligible for $n\gtrsim 40$. It also seems that a cusp will occur at $t=\tau$ in the limit $n \to \infty$  as is the case with the function $H'_1(t)$ computed from Eq. (\ref{EqHpBS}).
\begin{figure}[hbt]
\begin{center}
\includegraphics[width=8cm]{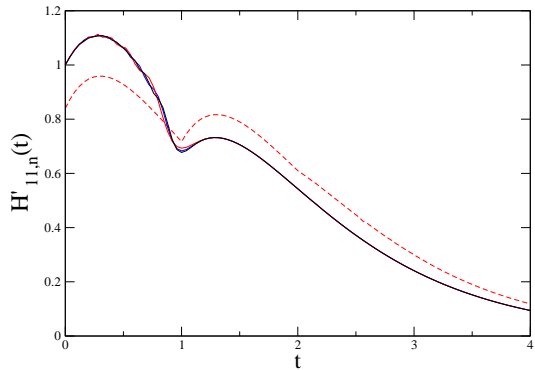}
 \caption{\label{Fig3} (Color on line)  The function $H'_{11,n}(t)$ for $n=20$ (solid red line), $n=30$ (solid blue line), and $n=40$ (solid black line), as obtained from the numerical inverse Fourier transform of Eq. (\ref{EqH1new}) for $\rho=0.5$, $a=2$, $c=4$, and $\tau =1$. The dashed red line represents $H'_{1}(t)$ computed from the numerical Fourier transform of Eq. (\ref{EqHpBS}). Note that this function is not equal to $1$ at $t=0$ (it is also nonzero for $t<0$).}
\end{center}
\end{figure}

\subsection{Influence of $\tau$ and $\rho$}

We are now in position to compare the TE rate estimated from Eq. (\ref{EqT21exact1}) with the  predictions of Eq. (\ref{EqSpec}) and investigate the influence of  the delay.
\begin{figure}[hbt]
\begin{center}
\includegraphics[width=8cm]{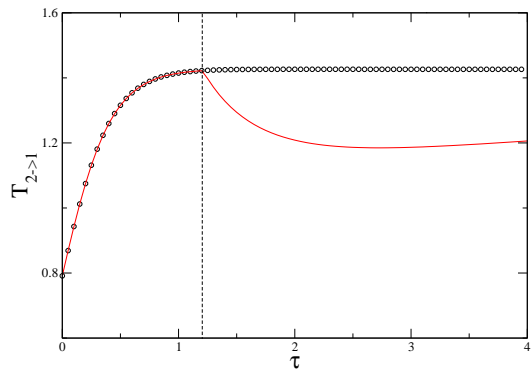}
 \caption{\label{Fig4} (Color on line)  Comparison between ${\cal T}_{2 \to 1}^{(n)}$ obtained from Eq. (\ref{EqT21exact1}) for $n=25$ (black circles) and the spectral formula (\ref{EqSpec}) of ${\cal T}_{2 \to 1}$  (red line) as a function of $\tau$ for $\rho=0.5$ and $c=4$.  Eq. (\ref{EqSpec}) is valid for $\tau<1.209$ only, as indicated by the vertical dashed line.}
\end{center}
\end{figure}

This is illustrated in Fig. \ref{Fig4}. Since $\rho c>b$ with our choice of the parameters, the discussion in the preceding subsection tells us that the spectral formula is valid up to $\tau=\tau^*\approx 1.209$. Indeed, we see in the figure that the agreement is excellent for $\tau \le \tau^*$ but that the two curves  deviate beyond $\tau^*$. The spectral formula then predicts a lower value of the TE rate, in line with the arguments of \cite{G1982}. More generally, our calculations show that ${\cal T}_{2 \to 1}$ is monotonically increasing with $\tau$ for $\rho c>0$, whereas it first decreases and then increases for $\rho c<0$. (It can be analytically shown that  $d/d\tau \ln {\cal T}_{2 \to 1}\vert_{\tau=0}=\rho c$.)  In both cases, ${\cal T}_{2 \to 1}$ goes to a finite value as $\tau \to \infty$. Note that the behavior of  ${\overline {\cal T}}_{2 \to 1}$  computed from Eq. (\ref{EqoverT21final}) is completely different. Consider for instance the simplest case of independent noises ($\rho=0$). Then ${\cal T}_{2 \to 1}=(\sqrt{b^2+c^2}-b)/2$ does not depend on $\tau$, as already noticed in \cite{BS2017}, whereas ${\overline {\cal T}}_{2 \to 1}$ decreases with $\tau$~\cite{note11}. (We chose not to plot ${\overline {\cal T}}_{2 \to 1}$ since it depends on the value of $a$, in contrast with ${\cal T}_{2 \to 1}$.)

\begin{figure}[hbt]
\begin{center}
\includegraphics[width=8cm]{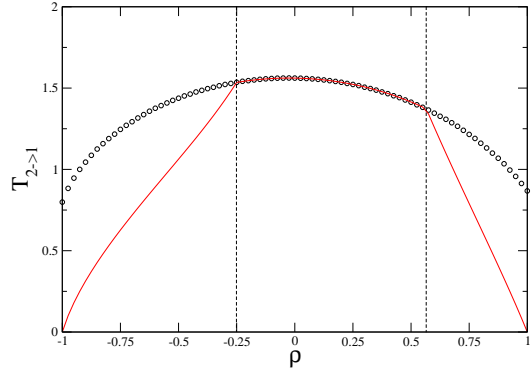}
 \caption{\label{Fig5} (Color on line) Same as Fig. \ref{Fig4} as a function of $\rho$ for $\tau=1$.  The spectral formula (\ref{EqSpec}) is only valid in the range of $\rho$ delimited by the vertical dashed lines.}
\end{center}
\end{figure}

To further illustrate the differences between Eqs.  (\ref{EqT21exact1}) and (\ref{EqSpec}), the behavior of ${\cal T}_{2 \to 1}$  as a function of $\rho$ for a fixed value of $\tau$ is shown in Fig. \ref{Fig5}. The spectral formula is now valid in the interval $-0.25\le \rho \le 0.565$, where the minimal value corresponds to $\rho=-b/c$ and the maximal value is the solution of the equation $\tau\sqrt{\rho^2c^2-b^2}-\arccos[-b/(\rho c)]=0$. The most striking feature is that ${\cal T}_{2 \to 1}$  goes to a finite value for $\rho=\pm1$,  at variance with the outcome of the spectral formula~\cite{note2}. It is a nontrivial fact that the TE rate remains finite even when the noises are strongly correlated or anti-correlated.

\subsection{Delay detection}

Finally, we discuss the issue of delay detection and estimation.  As pointed out in the introduction, this is a potentially important application of  transfer entropy, especially in neuroscience~\cite{WPPSSLL2013,BS2017}. Since it is still actively debated whether or not this method is reliable~\cite{CJJHKPP2017}, it is sensible to perform numerical tests on well-controlled dynamical systems, even as simple as the present one.

The  idea is that the finite-horizon TE, in the present case $T_{2 \to 1}(h)$, should display a maximum in the vicinity of $h=\tau$.  Indeed, as long as $h\lesssim\tau$, the trajectory of $X_2$ in the time interval $[t-\tau,t+h-\tau]$ provides a useful information about the future of $X_1$, which makes $T_{2 \to 1}(h)$ increase with $h$.  On the other hand, for $h\gtrsim\tau$, the trajectory of $X_2$ in the time interval $[t,t+h-\tau]$ is no longer taken into account  in $T_{2 \to 1}(h)$ since only the trajectory of $X_2$ prior to  $t$ contributes, by definition. Eventually, as $h\to \infty$, both $p(X_1(t+h)\vert \{{\bf X}(s)\}_{s\le t})$ and $p(X_1(t+h)\vert X_1^-(t))$ approach the stationary pdf $p(X_1)$ and $T_{2 \to 1}(h) \to 0$. 
\begin{figure}[hbt]
\begin{center}
\includegraphics[width=8cm]{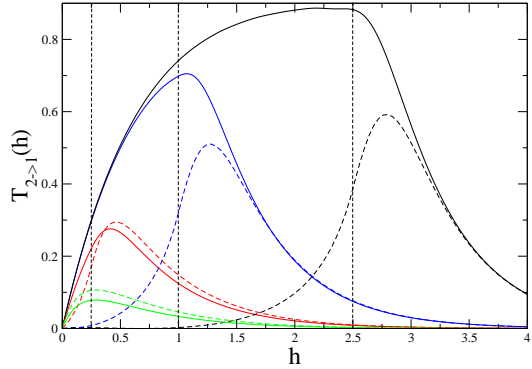}
 \caption{\label{Fig6} (Color on line) Finite-horizon TE  $T^{(n)}_{2 \to 1}(h)$ computed from Eq. (\ref{EqTn21}) for $n=30$, $\rho=0.5$ and different values of the delay: $\tau=2.5$ (black line),  $\tau=1$ (blue line), $\tau=0.25$ (red line), $\tau=0$ (green line). The model parameters are $a=2,b=1,c=4$. The dashed lines represent $\overline T_{2\to 1}(h)$.} 
\end{center}
\end{figure}

The above argument is only qualitative, though, and the accuracy of the estimate of $\tau$ must be checked numerically. Typical results are shown in Fig. \ref{Fig6} where the value of $\rho$ is arbitrarily fixed at $0.5$. Indeed, whereas the magnitude of $T_{2 \to 1}(h)$ depends on $\rho$, the overall behavior remains qualitatively unchanged, and in particular the position of maximum varies little~\cite{note4}.  In line with the qualitative argument above, we observe in the figure that the maximum in $T_{2 \to 1}(h)$ occurs just beyond $\tau$ for the two largest values of the delay (we recall that $b^{-1}$, the relaxation time of the process $X_2$, is here taken as the natural time scale in the problem). On the other hand, the agreement is not so good for the smallest values of $\tau$. The obvious problem is that $T_{2 \to 1}(h)$ exhibits a maximum at a certain time $h=h_0$ {\it even} when $\tau=0$. This time depends in a complicated way on the two relaxation times $a^{-1}$ and $b^{-1}$ and on the coupling strength $c$. It also differs from the time $h_{max}$ associated with the maximum of  the cross-correlation function $\phi_{21}(h)$ (for the case considered in Fig. \ref{Fig6}, $h_0\approx 0.29$ whereas $h_{max}\approx 0.15$). We may tentatively regard $h_0$  as the time it would take for $X_2$ to effectively influence $X_1$  if there were no inherent delay in the coupling between the two processes. Therefore, the lesson to be drawn  from the example in Fig. \ref{Fig6} is that $\tau$ must be significantly larger than $h_0$ to be properly estimated by scanning the horizon $h$ in $T_{2 \to 1}(h)$.  This is certainly a  limitation because the value of  $h_0$ is unknown in practice (e.g., in biochemical processes), although its order of magnitude may possibly be guessed.

\begin{figure}[hbt]
\begin{center}
\includegraphics[width=8cm]{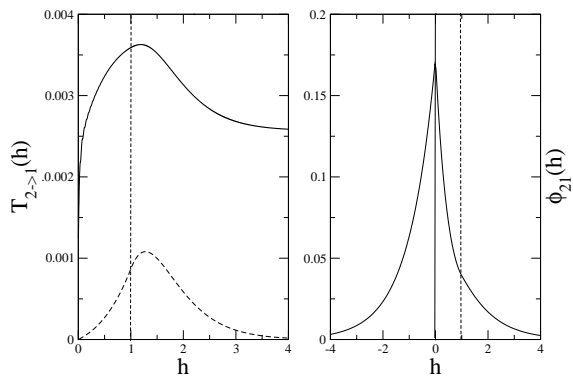}
 \caption{\label{Fig7} (Color on line) Left panel: Finite-horizon TE  $T^{(n)}_{2 \to 1}(h)$ for $n=30$, $\rho=0.5$, and $\tau=1$. The model parameters are $a=2,b=1,c=0.1$. The dashed line represents $\overline T_{2\to 1}(h)$. Right panel: Corresponding cross-correlation function $\phi_{21}(h)$.} 
\end{center}
\end{figure}
On the positive side, we wish to stress that $\tau$ can be detected via $T_{2 \to 1}(h)$ even when there is no clear signature of a delayed interaction in the cross-correlation function $\phi_{21}(h)$. (Otherwise, there would indeed be no profit in using $T_{2 \to 1}(h)$ which is much less easily extracted from time-series data than $\phi_{21}(h)$.) This occurs for instance when the coupling parameter $c$ is small, as illustrated in Fig. \ref{Fig7}. In this case, there is no maximum in $\phi_{12}(h)$, except in $h=0$, whereas the maximum in $T_{2 \to 1}(h)$ still takes place in the vicinity of $h=\tau$~\cite{note12}. 

We end this section with a short comment about ${\overline T}_{2\to 1}(h)$, which is also represented in Figs. \ref{Fig6} and \ref{Fig7} (this function is computed from Eq. (\ref{EqoverT12h}) with the labels $1$ and $2$ exchanged). Although the initial behavior of ${\overline T}_{2\to 1}(h)$ with $h$ differs from that of $T_{2\to 1}(h)$, in relation with the fact that the rates ${\overline {\cal T}}_{2\to 1}$ and ${\cal T}_{2\to 1}$ are quite different, ${\overline T}_{2\to 1}(h)$ also displays a maximum in the vicinity of  $h=\tau$ when $\tau$ is sufficiently larger than $h_0$.  This is interesting because this function can be more easily estimated from time-series data than $T_{2\to 1}(h)$ as it only requires the knowledge of the stationary correlation functions or the corresponding power spectral densities. One then circumvents the numerically challenging problem of estimating high-dimensional probability distributions (the so-called ``curse of dimensionality"~\cite{RHPK2012}).  Note however that the delay is better estimated with $T_{2\to 1}(h)$:  For instance in Fig. \ref{Fig6} and $\tau=2.5$, the maxima of $T_{2\to 1}(h)$ and ${\overline T}_{2\to 1}(h)$ are located at $h\approx 2.52$ and $h\approx 2.79$, respectively. In Fig. \ref{Fig7}, where $\tau=1$, the maxima are located at $1.19$ and $1.27$ respectively~\cite{note13}.

\section{Summary}

\label{Sec:Conclusion}

Is an information-theoretic measure such as the transfer entropy (TE) able to detect interaction delays in coupled systems?  This question, still debated, has prompted us to revisit the recent calculation performed in \cite{BS2017} for a linear stochastic process with a delayed coupling. By focusing on a simple model that can be solved analytically in continuous time, thus avoiding the difficulties arising from time discretization, one may hope to get a better understanding of the issue. However, even in the simple case of stationary Gaussian processes, the calculation of the finite-horizon TE (or equivalently Granger causality) in the presence of delay requires the solution of a nontrivial Wiener-Hopf factorization problem that was not properly treated in \cite{BS2017}. The main contribution of the present work is to provide an efficient solution to this problem in the case where the stochastic noises are correlated, as is often required in the modeling of real networks.  As a by-product, we have derived a compact expression of the zero-horizon TE rate. We have also clarified the conditions under which the spectral representation the TE rate is valid, an issue that seems to be  overlooked in the literature. Our numerical results for a bivariate model with unidirectional delayed coupling  show that the finite-horizon TE is indeed able to detect and estimate the delay (under some conditions, though), even when there is no clear signature in the cross-correlation function.  Interestingly, this is also true for the much simpler version of TE that  only takes into account the immediate past of the source and the target.

It is clear however that more analytical and numerical work remains to be done before reaching a comprehensive picture. A natural extension of the present work would be to consider multiple delays occurring in both directions (not to mention the case of multivariate systems). It would also be interesting to investigate the behavior of the TE in an oscillatory regime and in the vicinity of a Hopf bifurcation. We leave this to future investigations.

\appendix 
\label{Sec:Appendix}

\setcounter{figure}{0} \renewcommand{\thefigure}{A.\arabic{figure}} 

\renewcommand{\theequation}{A\arabic{equation}}

\section{Expressions of $\overline{T}_{i\to j}(h)$ and $\overline{{\cal T}}_{j\to i}$ in the presence of time delay}

In this appendix we derive the expressions of the simplified TE  $\overline{T}_{i\to j}(h)$ ($i,j=1,2$) defined by Eq. (\ref{EqTbarh}) and of the corresponding rate $\overline{{\cal T}}_{i\to j}$ for the bivariate process governed by Eq.   (\ref{EqDelay}). Our starting point is the expression of the conditional probability distribution function  $p({\bf x},t+h\vert {\bf x}',t)$ of a Gaussian stationary process in terms of the matrix $\Phi(t)$ of the correlation functions  $\phi_{ij}(h)\equiv \langle X_i(t)X_j(t+h)\rangle$, 
\begin{align} 
\label{EqPxxp}
p({\bf x},t+h\vert {\bf x}',t)=\frac{1}{2\pi\sqrt{\mbox{Det}\:{\boldsymbol \Sigma(h)}}}e^{-\frac{1}{2}[({\bf x}-{\bf G}(h){\bf x}')^T.{\boldsymbol \Sigma(h)}^{-1}.({\bf x}-{\bf G}(h){\bf x}')]}\, ,
\end{align}
where 
\begin{align} 
\label{EqGt}
{\bf G}(h)={\boldsymbol \Phi}(h)^T.{\boldsymbol \Sigma}^{-1}
\end{align} 
and 
\begin{align} 
\label{EqSigmat}
{\boldsymbol \Sigma}(h)={\boldsymbol \Sigma}-{\bf G}(h).{\boldsymbol \Phi}(h)\, .
\end{align}
We recall that $\Sigma=\Sigma(\infty)$ is the stationary covariance matrix with elements $\sigma_{ij}=\phi_{ij}(0)$ and  that 
\begin{align} 
p({\bf x})=\frac{1}{2\pi \sqrt{\mbox{Det}\:{\boldsymbol \Sigma}}}e^{-\frac{1}{2}({\bf x}^T.{\boldsymbol \Sigma}^{-1}.{\bf x})}\, .
\end{align}
One can check that the correlation functions are  indeed the second moments of $p({\bf x},t+h; {\bf x}',t)=p({\bf x},t+h\vert {\bf x}',t)p({\bf x}')$, i.e., $\phi_{ij}(h)=\int d{\bf x}\:d{\bf x}' \:x'_i  \:x_j \: p({\bf x},t+h; {\bf x}',t)$.

Consider for instance $\overline{T}_{1\to 2}(h)$. By integrating Eq. (\ref{EqPxxp}) over $x_1$ and then $p(x_2,t+h; {\bf x}',t)$ over $x'_1$, we successively obtain 
\begin{align} 
p(x_2,t+h\vert {\bf x}',t)&= \frac{1}{\sqrt{2\pi \sigma_{22}(h)}}\exp[-\frac{(x_2-G_{21}(h)x'_1-G_{22}(h)x'_2)^2}{2\sigma_{22}(h)}]\, ,
\end{align}
and 
\begin{align} 
p(x_2,t+h\vert x_2,t)&=\sqrt{\frac{\sigma_{22}}{2\pi [G_{21}^2(h)\mbox{Det}\:{\boldsymbol \Sigma}+\sigma_{22}(h)\sigma_{22}]}}\exp[-\frac{\sigma_{22}(x_2-\frac{G_{21}(h)\sigma_{12}+G_{22}(h)}{\sigma_{22}}x'_2)^2}{2[G_{21}^2(h)\mbox{Det}\:{\boldsymbol \Sigma}+\sigma_{22}(h)\sigma_{22}]}]\, .
\end{align}
This readily yields
\begin{align} 
\label{EqoverT12h}
\overline{T}_{1\to 2}(h)&\equiv\frac{1}{2}\int d{\bf x}\:d{\bf x}' \: \ln \frac{p(x_2,t+h\vert {\bf x}',t)}{p(x_2,t+h\vert x_2,t)}\nonumber\\
&=\frac{1}{2}\ln \left(1+\frac{G_{21}^2(h)\mbox{Det}\: {\boldsymbol \Sigma}}{\sigma_{22}(h)\sigma_{22}}\right)\nonumber\\
&=-\frac{1}{2}\ln \frac{\sigma_{22}}{\mbox{Det}\: {\boldsymbol \Sigma}}+\frac{1}{2}\ln \frac{\sigma_{22}^2-\phi_{22}^2(h)}{\sigma_{22}\mbox{Det}\: {\boldsymbol \Sigma}-\phi_{22}^2(h)\sigma_{11}+2\phi_{22}(h)\phi_{12}(h)\sigma_{12}-\phi_{12}^2(h)\sigma_{22}}\, .
\end{align}
Expanding $\phi_{12}(h)$ and $\phi_{22}(h)$ in powers of $h$, we then obtain the expression of the rate $\overline{{\cal T}}_{1\to 2}=\lim_{h \to 0^+}\overline{T}_{1\to 2}(h)/h$,
\begin{align} 
\label{EqoverT120}
\overline{{\cal T}}_{1\to 2}=-\frac{1}{4\sigma_{22}\mbox{Det}\: {\boldsymbol \Sigma}}\frac{[\dot \phi_{22}(0^+)\sigma_{12}-\dot \phi_{12}(0^+)\sigma_{22}]^2}{\dot \phi_{22}(0^+)}\, .
\end{align}
The corresponding expressions of $\overline{T}_{2\to 1}(h)$ and $\overline{{\cal T}}_{2\to 1}$ are obtained by exchanging the labels $1$ and $2$.

To proceed further and express $\overline{{\cal T}}_{1\to 2}$ and $\overline{{\cal T}}_{2\to 1}$ in terms of the $\sigma_{ij}$'s only, we need to compute the derivatives of the correlation functions at $t=0^+$. This can be done without fully solving the dynamics by using together the Fokker-Planck equation for the time-dependent probability distribution  $p({\bf x},t)$ and the differential equations satisfied by the $\phi_{ij}$'s. The Fokker-Planck equation is obtained as usual by starting from the definition $p({\bf x},t)=\langle \delta(X_1(t)-x_1)\delta(X_2(t)-x_2)\rangle$, inserting the Langevin equations, and using Novikov's theorem~\cite{N1964}. This yields  
\begin{align} 
\label{EqFP}
\frac{\partial p({\bf x},t)}{\partial t}&=\frac{\partial}{\partial x_1}[a_{11}x_1p({\bf x},t)+a_{12}\int dy\: y\:p({\bf x},t;y,t-\tau)]+\frac{\partial}{\partial x_2}[(a_{21}x_1+a_{22}x_2)p({\bf x},t)]\nonumber\\
&+D_{11}\frac{\partial^2}{\partial x_1^2}p({\bf x},t)+D_{22}\frac{\partial^2}{\partial x_2^2}p({\bf x},t)+2D_{12}\frac{\partial^2}{\partial x_1\partial x_2}p({\bf x},t)\,,
\end{align}
where $p({\bf x},t;y,t-\tau)=\langle \delta(X_1(t)-x_1)\delta(X_2(t)-x_2)\delta(X_2(t-\tau)-y)\rangle$ is a two-time probability density. (In passing, note that Eq. (\ref{EqFP}) is not a closed equation, which is a characteristic feature of time-delayed stochastic systems~\cite{GHL1999,F2005}.) Multiplying this equation by $x_1^2,x_2^2$ and $x_1x_2$, respectively, and integrating over ${\bf x}$, we obtain the following relations 
\begin{align} 
\label{Eqrel1}
a_{11}\sigma_{11}+a_{12}\phi_{21}(\tau)&=D_{11}\nonumber\\
a_{21}\sigma_{12}+a_{22}\sigma_{22}&=D_{22}\nonumber\\
(a_{11}+a_{22})\sigma_{12}+a_{21}\sigma_{11}+a_{12}\phi_{22}(\tau) &=2D_{12}\, .
\end{align}
On the other hand, from the differential equations for the correlation functions for $t\in [0^+,\tau]$,
\begin{align} 
\dot \phi_{11}(t)&=-a_{11}\phi_{11}(t)-a_{12}\phi_{21}(\tau-t)\nonumber\\
\dot \phi_{21}(t)&=-a_{11}\phi_{21}(t)-a_{12}\phi_{22}(\tau-t)\nonumber\\
\dot \phi_{12}(t)&=-a_{21}\phi_{11}(t)-a_{22}\phi_{12}(t)\nonumber\\
\dot \phi_{22}(t)&=-a_{21}\phi_{21}(t)-a_{22}\phi_{22}(t)\, ,
\end{align}
we obtain 
\begin{align} 
\label{Eqrel2}
\dot \phi_{11}(0^+)&=-a_{11}\sigma_{11}-a_{12}\phi_{21}(\tau)\nonumber\\
\dot \phi_{21}(0^+)&=-a_{11}\sigma_{12}-a_{12}\phi_{22}(\tau)\nonumber\\
\dot \phi_{12}(0^+)&=-a_{21}\sigma_{11}-a_{22}\sigma_{12}\nonumber\\
\dot \phi_{22}(0^+)&=-a_{21}\sigma_{12}-a_{22}\sigma_{22}\, .
\end{align}
Combining Eqs. (\ref{Eqrel1}) and (\ref{Eqrel2}) then gives
\begin{align} 
\dot \phi_{11}(0^+)&=-D_{11}\nonumber\\
\dot \phi_{21}(0^+)&=-2D_{12}+a_{21}\sigma_{11}+a_{22}\sigma_{12}\nonumber\\
\dot \phi_{12}(0^+)&=-a_{21}\sigma_{11}-a_{22}\sigma_{12}\nonumber\\
\dot \phi_{22}(0^+)&=-D_{22}\, .
\end{align}
Inserting these expressions into Eq. (\ref{EqoverT120}) and into the corresponding equation for $\overline{{\cal T}}_{2\to 1}$, we finally obtain
\begin{align} 
\label{EqoverT12final}
\overline{{\cal T}}_{1\to 2}=\frac{a_{21}^2\mbox{Det} \:{\boldsymbol \Sigma}}{4D_{22}\sigma_{22}}
\end{align}
and
\begin{align} 
\label{EqoverT21final0}
\overline{{\cal T}}_{2\to 1}=\frac{[D_{11}\sigma_{12}+(a_{21}\sigma_{11}+a_{22}\sigma_{12}-2D_{12})\sigma_{11}]^2}{4D_{11}\sigma_{11}\mbox{Det}\: {\boldsymbol \Sigma}}\, .
\end{align}
Note that $\overline{{\cal T}}_{1\to 2}$  depends on $\tau$ and $\rho$ only through the covariances $\sigma_{ij}$'s, so that Eq. (\ref{EqoverT12final})  is formally the same equation as the one derived in \cite{IS2015} for a simple bipartite Ornstein-Uhlenbeck process. 

Finally, for the model studied in section \ref{subsec:T21}, Eq. (\ref{EqoverT21final0})  reduces to 
\begin{align} 
\label{EqoverT21final}
\overline{{\cal T}}_{2\to 1}=\frac{[\frac{1}{2}\sigma_{12}+(b\sigma_{12}-\rho)\sigma_{11}]^2}{2\sigma_{11}\mbox{Det}\: {\boldsymbol \Sigma}}\, ,
\end{align}
with
\begin{align} 
\sigma_{11}&=\frac{2bc\rho e^{-a\tau}+ab+b^2+c^2}{2ab(a+b)}\nonumber\\
\sigma_{12}&=\frac{ce^{-b\tau}+2b\rho}{2b(a+b)}\nonumber\\
\sigma_{22}&=\frac{1}{2b}\, .
\end{align}

\setcounter{figure}{0} \renewcommand{\thefigure}{B.\arabic{figure}} 

\renewcommand{\theequation}{B\arabic{equation}} 

\section{Gamma-distributed delay}

An approximation often used in the context of biological modeling~\cite{M1989,C2010,H2011} consists in replacing the discrete delay kernel $\delta(t-\tau)$ in the time domain by a sequence of gamma distributions $\delta(t-\tau)\approx g_n(t,\tau/n)$ where $g_n(t,T)= [(n-1)!\: T^n]^{-1}t^{n-1}e^{-t/T}$.  For instance, at the lowest order $n=1$, the memory kernel reduces to a low pass filter with bandwidth $\tau^{-1}$ and $X_2(t-\tau)$ in  Eq. (\ref{EqDelay}) is replaced by $(1/\tau)\int_{-\infty}^t ds \: e^{-\frac{t-s}{\tau}}X_2(s)$.  In the frequency domain, this approximation amounts to replacing  $e^{i\omega \tau}$ by $(1-i\omega \tau/n)^{-n}$. Eq. (\ref{EqS11new}) in the main text is then replaced by 
\begin{align} 
\label{EqS11new2}
S_{11,n}(\omega)=\frac{{\cal P}_n(\omega)}{(a^2+\omega^2)(b^2+\omega^2)(1+\frac{\omega^2\tau^2}{n^2})^n}\, ,
\end{align}
where 
\begin{align} 
\label{EqPnnew2}
{\cal P}_n(\omega)=(\omega^2+b^2+c^2)(1+\frac{\omega^2\tau^2}{n^2})^n+\rho c [(1+i\frac{\omega \tau}{n})^{n}(b+i\omega)+(1-i\frac{\omega \tau}{n})^{n}(b-i\omega)]\, , 
\end{align}
and the Wiener-Hopf causal factor is
\begin{align}
\label{EqH11pnew2} 
H'_{11,n}(\omega)= \frac{i\prod_{k=1}^{n+1}(\omega-\omega_k)}{(\omega+ia)(\omega+ib)(\omega+in/\tau)^{n}}\, .
\end{align}
Noting that $\dot H'_{12,n}(0^+)=0$ with this approximation of $e^{i\omega \tau}$, we finally arrive at
\begin{align} 
\label{EqT21exact2}
{\cal T}_{2 \to 1}^{(n)} &=\frac{1}{2} [a-\frac{n}{\tau}-\lim_{\omega \to \infty} \omega^2\big(H'_{11,n}(\omega)-\frac{1}{n/\tau-i\omega}\big)]\nonumber\\
&=\frac{1}{2}[-b+i\sum_{k=1}^{n+1}\omega_k-\frac{n^2}{\tau}]\, ,
\end{align}
which replaces Eq. (\ref{EqT21exact1}).
\begin{figure}[hbt]
\begin{center}
\includegraphics[width=8cm]{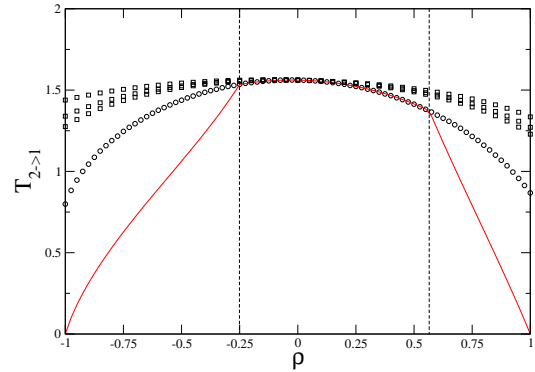}
 \caption{\label{Fig8} (Color on line) Same as Fig. 5 in the main text with in addition the predictions of Eq. (\ref{EqT21exact2}) for $n=30,60,90$ (black squares, from top to bottom).}
\end{center}
\end{figure}  

The advantage of this representation of $\delta(t-\tau)$ is that  a finite value of $n$ may provide a good description of a given physical or biological process (whereas taking $n$ finite in the Laguerre shift formula (\ref{EqLaguerre}) does not correspond to a {\it bona fide}  Langevin process with distributed delay). Eqs. (\ref{EqS11new2})-(\ref{EqH11pnew2}) then give the exact solution of the corresponding Wiener-Hopf factorization and in turn the exact expression of the TE.   On the other hand, as illustrated in Fig. \ref{Fig8}, the convergence with $n$ is very slow. Therefore, this approximation is not appropriate for dealing with a true discrete delay.


\begin{thebibliography}{10}
\bibitem{note00} If ${\bf X}$ is a Markov process, one has $p(X_j(t+h)\vert {\bf X}^-(t))=p(X_j(t+h)\vert {\bf X}(t))$ and thus ${\overline T}_{i \to j}(h)-T_{i \to j}(h)=H[X_j(t)\vert X_j(t)] -H[X_j(t)\vert X_j^-(t)]$ where $H[.\vert .]$ is the conditional Shannon entropy. This difference is always non-negative.
\bibitem{note5} The continuous population approximation is generally valid in signaling cascades where molecular populations are large. On the other hand, the linearization of the Langevin equations is an approximation whose validity depends on the details of the underlying biochemical reaction network and which must be checked on a case-by-case basis.
\bibitem{note10} The polynomial $[1+\omega^2\tau^2/(4n^2)]^{n}{\cal P}_n(\omega)$ can be rewritten as $(AB)^n{\cal P}_n(\omega)=[\omega (AB)^n+i(\rho c/2)(A^{2n}-B^{2n})]^2+[b(AB)^n+(\rho c/2)(A^{2n}+B^{2n})]^2+(1-\rho^2)c^2(AB)^{2n}$, where $A=1+i\omega \tau/(2n)$ and $B=A^*=1-i\omega \tau/(2n)$. If $\omega$ is real, $A^{2n}+B^{2n}$ and $A^{2}-B^{2n}$ are real and imaginary quantities, respectively, and $(AA^*)^{n}{\cal P}_n(\omega)$ is then a sum of three positive terms. Therefore ${\cal P}_n(\omega)$ has no real roots.
\bibitem{note1} The evolution of the roots of the polynomial ${\cal P}_n(\omega)$ with $n$ is indeed quite complicated. In particular, some of the roots  are ``spurious"  in the sense that they are not zeros of the function $\omega^2+b^2+c^2+2\rho c \:v(\omega,b,\tau)$ (i.e., the  actual zeros of $S_{11}(\omega)$) in the limit $n \to \infty$. However,  they must be included in the calculation to ensure that $i\sum_{k=1}^{n+1}\omega_k-2n^2/\tau$ goes asymptotically to a finite value.
\bibitem{note11} This can be easily understood. Since $X_1(t+h)$ for $h$ infinitesimal is determined by $X_1(t)$ and $X_2(t-\tau)$, the additional information provided by the knowledge of $X_2(t)$ becomes less and less relevant as $\tau$ increases, which makes ${\overline {\cal T}}_{2\to 1}$ decrease. On the other hand, if the whole past of $X_1$ up to $t$ is already known and one adds information about the whole past of $X_2$ (that includes its value at $t-\tau$), then $\tau$ no longer plays a role and ${\cal T}_{2\to 1}$ remains constant. For $\rho\ne 0$, however, this reasoning is no longer valid.
\bibitem{note12} One could object that there is a kink at $h=\tau$ in both $\phi_{11}(h)$ and $\phi_{21}(h)$ which signals the presence of a delay in the interaction. However, it is very difficult in general to recognize such a feature in experimental data.
\end{thebibliography}
\end{document}